\documentclass[nofootinbib,twocolumn,aps,pre,superscriptaddress,citeautoscript,floatfix]{revtex4-1}
\usepackage{graphics}
\usepackage{graphicx}
\usepackage{mathrsfs}

\begin{document}

\title{Lamellar Diblock Copolymers on Rough Substrates: Self-consistent Field Theory Studies}
\author{Xingkun Man$^{*}$}
\affiliation{Center of Soft Matter Physics and its Applications, Beihang University, Beijing 100191, China}
\affiliation{School of Physics and Nuclear Energy Engineering, Beihang University, Beijing 100191, China}
\author{Jiuzhou Tang}
\affiliation{Beijing National Laboratory Molecular Sciences, Joint Laboratory of Polymer Science and Materials,
Institute of Chemistry, Chinese Academy of Sciences, Beijing 100190, China}
\author{Pan Zhou}
\affiliation{Department of Physics, Beijing Normal University, Beijing 100875, China}
\author{Dadong Yan}
\affiliation{Department of Physics, Beijing Normal University, Beijing 100875, China}
\author{David Andelman$^{*,}$}
\affiliation{Raymond and Beverly Sackler School of Physics and Astronomy, Tel Aviv University, Ramat Aviv 69978, Tel Aviv, Israel}
\affiliation{Kavli Institute for Theoretical Physics China, CAS,
Beijing 100190, China}

\begin{abstract}

We present numerical calculations of lamellar phases of di-block copolymers (BCP) confined between two
surfaces, where the top surface is flat and the bottom one is corrugated. The corrugated substrate is assumed
to have a single $q$-mode of lateral undulations with a wavenumber $q_s$ and amplitude $R$. We focus on the
effects of substrate roughness, parameterized by the dimensionless quantity, $q_sR$, on the relative stability
between parallel and perpendicular orientations of the lamellar phase. The competition between film
confinement, energy cost of elastic deformation and gain in surface energy induces a parallel-to-perpendicular
transition of the BCP lamellae. Employing self-consistent field theory (SCFT), we study the critical value,
$(q_sR)^*$, corresponding to this transition. The $(q_sR)^*$ value increases as function of the surface
preference towards one of the two BCP components, and as function of film thickness. But, $(q_sR)^*$ decreases
with increasing values of the Flory-Huggins parameter, $N\chi_{AB}$. Our findings are equivalent to stating
that the critical $(q_sR)^*$ value decreases as the BCP molecular weight or the natural BCP periodicity increases. We further show that the
rough substrate can overcome the formation of parallel lamellae in cases where the top surface has a preference
towards one of the two BCP components. Our results are in good agreement with previous experiments, and
highlight the physical conditions behind the perpendicular orientation of lamellar phases, as is desired in
nanolithography and other industrial applications.

\end{abstract}

\maketitle

\section{Introduction}

Block copolymers (BCP) are polymer systems where each of their chains is composed of two or more
chemically-distinct homopolymer blocks, covalently tethered together. As a result, BCP systems can
spontaneously self-assemble at thermodynamical equilibrium into exquisitely ordered nano-structures
\cite{Glenn06}. The phase behavior of di-BCP melts, where each linear chain is composed of two blocks, (denoted
hereafter as A and B), has been studied extensively in recent decades, and shows a rich variety of
three-dimensional morphologies including lamellae, hexagonally close-packed cylinders, BCC packing of spheres,
and gyroid networks \cite{Matsen94,Bates94}. The characteristic length scale in these well-defined structural
phases ranges from a few nanometers to hundreds of nanometers, and can offer an attractive alternative to patterning technology \cite{Kim10,Bates90}. Besides applications in nano-lithography, BCP films may offer novel opportunities in more traditional applications such as adhesive, hydrophobic and anti-reflective surfaces, as well as in the textile industry \cite{Sinturel13}.

Most of the BCP applications rely on casting them as thin films since this is the most appropriate form to
construct a surface pattern that can later be transferred onto a substrate, with potential applications as
functional nanoscale devices \cite{Hamley09}. A perpendicular orientation of BCP lamellae or cylinders, with
respect to the underlying substrate, is usually desirable for most material and engineering applications
\cite{Bates14}. During recent decades, various techniques have been developed to obtain such perpendicular
lamellae or cylinders, including nonpreferential (neutral) interfaces \cite{Kim10,Liu09}, topographically
varying substrates \cite{Park09a} or top surfaces \cite{Man11,Man12}, variations in polymeric block
architecture \cite{Khanna06,Matsen10} and film thickness, as well as solvent annealing \cite{Sinturel13}.

It is also possible to use corrugated substrates to obtain perpendicular BCP lamellae or cylinders. Sivaniah
\emph{et al} \cite{Sivaniah03,Sivaniah05} reported the effect of substrate roughness on the orientation of
lamellae of symmetric poly(styrene)-block-poly(methyl methacrylate) (PS-b-PMMA). They identified a critical
substrate roughness, $(q_{\rm{s}}R)^{\rm{*}}$, above which a perpendicular orientation was observed, where
$q_{\rm s}$ and $R$ are the lateral wavenumber and its amplitude, respectively. They also found that the value
of $(q_{\rm{s}}R)^{\rm{*}}$ varies with BCP molecular weight (or the periodicity of BCP lamellae). In a more
recent study, Kulkarni \emph{et al} \cite{Kulkarni12} extended the results to include fractal substrate
topography. A high fractal dimension of the rough substrate, in conjunction with an optimal surface
energy of PS-b-PMMA in contact with the substrate, results in a complete perpendicular orientation of lamellar
micro-domains.

In a separate work, Kim \emph{et al} \cite{Char13} investigated a film of PS-b-PMMA placed on an ordered
nanoparticle (NP) monolayer. The substrate roughness is described by the parameter $q_{\rm{sub}}r$, where
$q_{\rm{sub}}$ and $r$ are the wavenumber of substrate roughness and the radius of NP, respectively. A
transition from parallel to perpendicular orientation of BCP lamellae or cylinders has been found by increasing
the value of $q_{\rm{sub}}r$. Furthermore, it was shown that the orientation of thin films of BCP is strongly
influenced by the film thickness. This is due to the commensurability matching between film thickness and
domain spacing.


In addition to the experimental situation, there are few theoretical works addressing the self-assembly of
BCP films on corrugated surfaces. Turner and Joanny~\cite{Turner1992}, and Tsori \emph{et al}~\cite{Yoav03,Yoav05}
used the analogy between smectic
liquid crystals and lamellar BCP, and compared the phenomenological free energy of parallel and perpendicular
lamellae on corrugated substrates.  In Refs.~\cite{Yoav03,Yoav05} it was shown
that for a fixed corrugation periodicity, the perpendicular
orientation is preferred for large corrugation amplitude and/or large lamellae periodicity. Moreover, for a
fixed BCP natural periodicity, the perpendicular orientation is preferred for surfaces having large
corrugation amplitude at short wavelengths.

Motivated by previous experimental works \cite{Kulkarni12}, Ranjan \emph{et al} \cite{Ranjan12} conducted a
scaling analysis of a single BCP lamella on fractal surfaces, which gives additional evidence that
the substrate fractal dimension is an important factor in directing the orientation of BCP lamellae. Even more
recently, Ye \emph{et al} \cite{Ye14} studied morphological properties of lamellae-forming di-BCPs on
substrates with square-wave grating patterns by using self-consistent field theory (SCFT). They found three
possible lamellar orientations with respect to the substrate and trench direction, but without addressing
the key factors that determine the critical substrate roughness at the parallel-to-perpendicular phase-transition of BCP micro-domains.

We note that these previous studies have provided insight on how substrate roughness affects the relative
stability of parallel and perpendicular BCP micro-domains on non-flat surfaces. However, to date, systematic
studies addressing the combined effect of substrate corrugation amplitude and lateral wavenumber, film
thickness and BCP periodicity on the domain orientation of BCP films are still missing. In this paper, we
present a comprehensive and detailed SCFT study of di-BCP films constrained between a top flat surface and a bottom corrugated substrate. Our aim is to investigate the role played by substrate geometry, relative surface preference of the two BCP components, and BCP film properties, including film thickness, the Flory-Huggins parameter, $N\chi_{\rm{AB}}$, between the two monomers, and the lamellar periodicity, on the parallel-to-perpendicular phase-transition.

In the next section, we introduced the SCFT formalism, and our numerical scheme. In Sec. III, the corrugated surface and BCP film design are presented, while in Sec. IV, we show the calculated phase diagrams of BCP lamellae on corrugated substrates. Discussion of our results and comparison with previous models and experiments are presented in Sec. V, followed by a summary and conclusions.

\section{Theoretical Framework}

\subsection{The SCFT Scheme}

We use self-consistent field theory (SCFT) to investigate the lamellar phase of A/B di-BCP confined between two
surfaces, where the top surface is flat and the bottom one is corrugated. We consider a melt of $n_c$ chains,
each composed of $N=N_A+N_B$ monomers. For simplicity, the Kuhn length, $b$, is assumed to be the same for the
A and B monomers, yielding an equality between the molar fraction and the volume one.
The A-component molar fraction is $f=N_A/N$ and that of the B-component is $1-f$. The BCP
film has a total volume $\Omega$, lateral area $\cal{A}$, and thickness $L=\Omega/\cal{A}$. Hereafter, we
concentrate on symmetric di-BCP, i.e., $f=0.5$.

In order to facilitate the numerical convergence, it is convenient
to replace the sharp interface between the BCP film and the hard bounding surfaces
by a ``softer" wall  with a smeared interface having a small width. This is done by
introducing an artificial third (wall) component~\cite{Matsen97, Bosse07, Glenn06}.
The local incompressibility condition is
$\phi_A(\textbf{r})+\phi_B(\textbf{r})+\phi_w(\textbf{r})=1$,
where $\phi_A$, $\phi_B$ and $\phi_w$ are the A, B and wall volume fractions within our simulation box,
respectively. This condition is replaced with a compressible
one, by adding an energetic penalty cost for local density deviations from the incompressibility condition.
The penalty term is written as:
\begin{equation}
\zeta\Bigl(\phi_A(\textbf{r})+\phi_B(\textbf{r})+\phi_w(\textbf{r})-1\Bigr)^2 \, ,
\end{equation}
and has a magnitude controlled by an ``energy'' parameter $\zeta$ (in units of $k_{B}T$, where $k_B$ is the Boltzmann
constant and $T$ is the temperature).

The direction parallel to the substrate is
chosen to be along the $x$-direction, and the perpendicular one is in the $y$-direction. The system is assumed
to be translational invariant in the third $z$-direction, which means that the numerical calculations are
performed in a two-dimensional (2d) box. All lengths, hereafter, are rescaled with the chain radius of
gyration, $R_g=\sqrt{Nb^2/6}$. With these conventions, the Hamiltonian of the BCP film confined between two
surfaces can be expressed as a functional of two local fields: a pressure field $W_{+}(\bf{r})$ and an exchange
potential field $W_{-}(\bf{r})$~\cite{Glenn06},

\begin{eqnarray}\label{f1}
H[W_{+},W_{-}] & = & C\int d^2\textbf{r}\left(\frac{\left[W_{-}(\textbf{r})\right]^2}
{N\chi_{\rm{AB}}} - \frac{2Nu}{N\chi_{\rm{AB}}}\phi_{w}(\textbf{r})W_{-}(\textbf{r})
\right. \nonumber\\
 &+& \left. \frac{ \left[W_{+}(\textbf{r})\right]^2 - 2\zeta N\phi(\textbf{r})
  iW_{+}(\textbf{r})}{N\chi_{\rm{AB}} + 2N\zeta}\right) \nonumber\\
  & -&  C \Omega\bar{\phi}\ln Q[W_A,W_B],
\end{eqnarray}
where $C=\rho_0R_g^3/N$ is a normalization factor, $\rho_0=(Nn_c+N_w)/\Omega$ is the total number density,
$\Omega$ is the entire volume (including the walls) of the simulation box,
$N_w$ is the total number of ``wall monomers", and $\phi(\textbf{r})=\phi_A(\textbf{r})+\phi_B(\textbf{r})$ is
the dimensionless volume fraction of the polymer. The Flory-Huggins parameter between the A and B monomers is
$\chi_{\rm{AB}}$, and $u=\chi_{w\rm{A}}-\chi_{w\rm{B}}$ is the relative interaction between the wall and the
A/B components, where $\chi_{w\rm{A}}$ and $\chi_{w\rm{B}}$ are the interaction parameters between the wall (as
a 3rd component) and the A or B components, respectively. For example, a positive $u>0$ means that the surface
prefers the A component. For simplicity, hereafter we absorb the factor of $N$ into the definition of  $\zeta$ and $u$: $N\zeta \to \zeta$ and $Nu \to u$.

The functional $Q[W_A,W_B]=\Omega^{-1}\int d^2\,\textbf{r}~q(\textbf{r},s{=}1)$ is the single-chain partition
function in the presence of the two conjugate fields, $W_A(\textbf{r})=iW_{+}(\textbf{r})-W_{-}(\textbf{r})$
and $W_B(\textbf{r})=iW_{+}(\textbf{r})+W_{-}(\textbf{r})$, where the propagator $q(\textbf{r},s)$ is the
solution of the modified diffusion equation,
\begin{equation}\label{f2}
\frac{\partial q(\textbf{r},s)}{\partial s}=\nabla ^2q(\textbf{r},s) - W(\textbf{r},s)q(\textbf{r},s),
\end{equation}
satisfying the initial condition $q(\textbf{r},s{=}0)=1$. $W(\textbf{r})=W_A(\textbf{r})$ for $0\leq s< f$
and $W(\textbf{r})=W_B(\textbf{r})$ for $f\leq s\leq 1$, where $s$ is the curvilinear coordinate along the
A/B chain contour. Finally, $\bar{\phi}={\Omega}^{-1}\int d^2\textbf{r}\, \phi(\textbf{r})$ is the
polymer volume fraction averaged over $\Omega$.

Within the mean-field approximation, we can obtain the thermodynamic properties of the confined
BCP film from a variational principle of the Hamiltonian in Eq.~(\ref{f1}),
\begin{eqnarray}\label{f4}
  \frac{\delta H[W_{+},W_{-}]}{\delta (iW_{+}(\textbf{r}))}&=&0 \, ,\nonumber\\
    & & \nonumber \\
  \frac{\delta H[W_{+},W_{-}]}{\delta W_{-}(\textbf{r})}&=&0 \, .
\end{eqnarray}
This means that
\begin{eqnarray}\label{f5}
  \frac{\delta H}{\delta (iW_{+})}&=&C\left[\phi_A(\textbf{r})+\phi_B(\textbf{r})-
  \frac{2\zeta\phi(\textbf{r})+2iW_{+}}{N\chi_{\rm{AB}}+2\zeta }\right] \nonumber \, \\
  & & \nonumber \\  & & \nonumber \\
  \frac{\delta H}{\delta W_{-}}&=&
  C\left[-\phi_A(\textbf{r})+\phi_B(\textbf{r})-\frac{2u\phi_w(\textbf{r})-2W_{-}}{N\chi_{\rm{AB}}}\right]\, .\nonumber\\
  & &
\end{eqnarray}
Equation~(\ref{f4}) can be solved numerically by solving the modified diffusion equation, Eq.~(\ref{f2}),
with spatially-periodic boundary conditions and the following temporal relaxation equations,
\begin{eqnarray}
\frac{\partial (iW_{+}(\textbf{r},t))}{\partial t}&=&\frac{\delta H\left[W_{+},W_{-}\right]}{\delta (iW_{+})} \, ,\nonumber\\
  & & \nonumber \\
\frac{\partial W_{-}(\textbf{r},t)}{\partial t}&=&-\frac{\delta H\left[W_{+},W_{-}\right]}{\delta W_{-}}\, .
\end{eqnarray}
We use the well-documented pseudo-spectral method with FFTW to solve the modified diffusion equation, and an explicit Euler
scheme (1st order in the iteration time) to update the field configurations to their saddle
points. A detailed formulation of the numerical procedures for the SCFT model and their implementation to BCP systems can be found elsewhere
\cite{Bosse07, Hur09, Takahashi12}.

\begin{figure*}[h!t]
\begin{center}
{\includegraphics[bb=0 0 350 268, scale=0.65,draft=false]{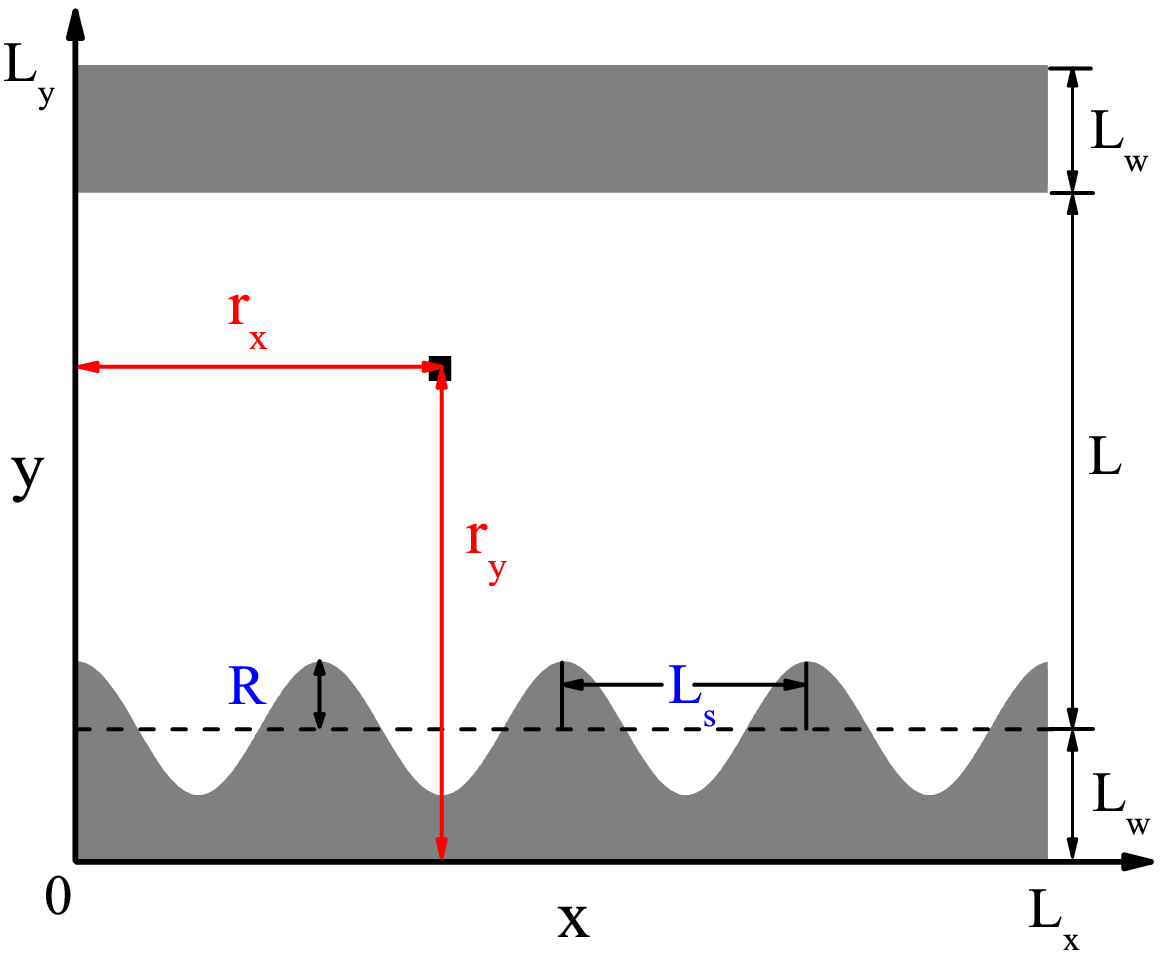}}
\caption{
\textsf{Schematic illustration of a BCP film confined between two surfaces. The two-dimensional calculation box
has the size $L_x\times L_y$, where $L=L_y-2L_w$ is the averaged BCP film thickness, and $L_w$ is the wall thickness
(see text). The corrugated substrate is described by a height function: $h(x)=R\cos
(q_s x)$, with lateral wavenumber $q_s$ and amplitude $R$. Any
point residing inside the film, $\textbf{r}=(r_{x}, r_{x})$, is parameterized by its $\rm{x}$-axis and $\rm{y}$-axis coordinates.
The coordinate origin is taken at the bottom left corner of the simulation box. All lengths are rescaled by the chain radius of gyration, $R_g$.}}
\end{center}
\end{figure*}

\subsection{Corrugated Substrate Design}

Our simulation setup is shown schematically in Fig.~1, where the 2d simulation box of size $L_x\times L_y$
includes the BCP
film and the solid boundaries, and the average wall thickness is
$L_w$. Hence, the average BCP film thickness $L$ is given by $L=L_y - 2L_w$. The top surface is flat,
while the bottom surface is corrugated, and is described by a
height function

\begin{equation}
h(x)=R\cos(q_s x)\, ,
\end{equation}
having a single $q$-mode with wavenumber $q_s$ and amplitude $R$.

The wall density, $\phi_{w}(\textbf{r})$, has a pre-assigned shape that is fixed (frozen) during the iterations.
The top flat wall is modelled as a rectangle of size $L_x \times L_w$ (Fig.~1), and characterized by a
smoothly-varying wall function:
\begin{equation}
\phi_{w}(\textbf{r})= \frac{1}{2}+\frac{1}{2}\tanh\Bigl[\frac{r_y - L_{w}-L}{\delta}\Bigr],
\label{1a}
\end{equation}
where $\delta$ parameterizes the interface width, and $r_y$ is the distance
from the bottom box boundary. For the bottom corrugated surface, we impose a similar smoothly-varying wall function
\begin{equation}
\phi_{w}(\textbf{r})= \frac{1}{2}-\frac{1}{2}\tanh\Bigl[\frac{r_y - R\cos (q_s r_x)-L_{w}}{\delta}\Bigr],
\end{equation}
where $r_x$ is the distance to the left box boundary. Such a definition means that $\phi_{w}(\textbf{r})=1$
inside the wall region of the rectangular box, and $\phi_{w}(\textbf{r})=0$ inside the BCP film.
However, it generates a ``soft" interface that is characterized by a narrow and smooth transition region of thickness
$\delta$. Hereafter, we set $\delta =0.1$ and $\zeta=1000$ for all simulations, following previous simulation
work~\cite{Takahashi12}. Our results are not sensitive for values of $\delta\le 0.1$, and consequently
$\delta=0.1$ is chosen for converge purposes. The value $\zeta=1000$ is large enough to model an incompressible system, yet facilitate numerical converge.

\section{Results}

\subsection{The Parallel and Perpendicular Lamellar Orientations} 

Symmetric BCPs yield thermodynamically stable lamellar phases with a bulk periodicity, $L_0$.
The immediate effect of the corrugation can be seen in Fig.~2.
When the substrate roughness is large enough, $q_sR\simeq 0.44$ (Fig.~2(a)), the
surface-induce distortion only propagates up to the second layer and all other lamellae are unperturbed.
However, for small substrate roughness, $q_sR\simeq 0.09$ as in Fig.~2(b), the lamellae follow the surface contour, and the distortion are longer range. It is important to note that the results are obtained for a finite film thickness where the top surface is flat and neutral ($u_{\rm top}=0$). These results have smaller penetration length as compared with previous scaling~\cite{Turner1992,Yoav03,Yoav05} for infinite stacks of lamellae, where the penetration length scales as $\sim q_0/q_s^2\sim L_s^2/L_0$.

\begin{figure*}[h!t]
\begin{center}
{\includegraphics[bb=0 23 418 190, scale=0.9,draft=false]{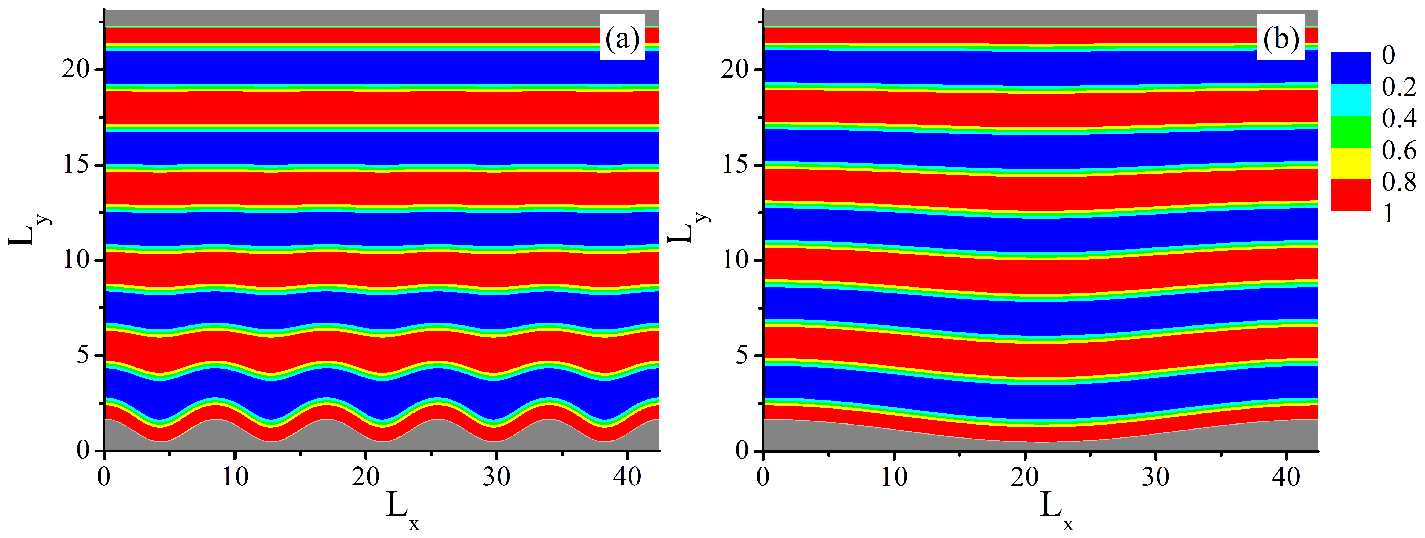}}
\caption{
\textsf{Parallel lamellar phase on a corrugated substrate.
(a) The lateral wavelength $q_s=5(2\pi/L_x)$ and amplitude $R=0.6$, resulting in a surface roughness, $q_sR\simeq 0.44$. In this case, the surface-induced distortion penetrates only through the second layer. (b)
A much smaller $q_sR\simeq 0.09$,  resulting from $q_s=2\pi/L_x$ and $R=0.6$. Here the lamellar first four layers follow the surface topography. The color code corresponds to five discrete intervals of local monomer density $0\leq\phi_A(\textbf{r}) \leq 1$, as is depicted in (b). Other parameters are: $L_x=42.5$, $L=21.25$, $R=0.6$, $u=3.3$, $u_{\rm{top}}=0$, and $N\chi_{\rm{AB}}=25$.}
}
\end{center}
\end{figure*}

Figure 3 shows examples of parallel and perpendicular lamellar phases in contact with a corrugated substrate in the strong segregation regime, $N\chi_{\rm{AB}}=25$. The bottom substrate in (a) and (b), having a moderate roughness, $q_sR\simeq0.25$, is attractive to the A component (marked in red) with $u=3.05$, while the top surface is neutral ($u_{\rm top}=0$). Clearly, the perpendicular lamellar phase (L$_\perp$) in (a) is almost unperturbed, as compared to its bulk phase. However, the parallel one (L$_\parallel$) in (b) adjusts its shape due to the surface corrugation. This is a result of the competition between the cost of BCP elastic deformation close to the corrugated surface and the gain in its surface energy. The results both in (a) and (b) agree well with previous analytical results carried out by Tsori {\it et al} \cite{Yoav03,Yoav05}.

For the case of large substrate roughness, $q_sR\geq 1$, the lamellae are strongly deformed, and it is hard to recognize whether their equilibrium structure is an L$_{\perp}$ or L$_{\parallel}$ phase. Such pronounced deformations are shown in Fig.~3(c) and (d), where the bottom substrate roughness is $q_sR\simeq1.23$, while all other conditions are the same as in (a) and (b). To be able to have meaningful predictions, we hereafter consider only substrates that have moderate roughness, $q_sR<1$, and surface preference $u$ that is less than the interaction between the A and B components, $u<N\chi_{AB}$.

\begin{figure*}[h!t]
\begin{center}
{\includegraphics[bb=0 0 380 282, scale=1.00,draft=false]{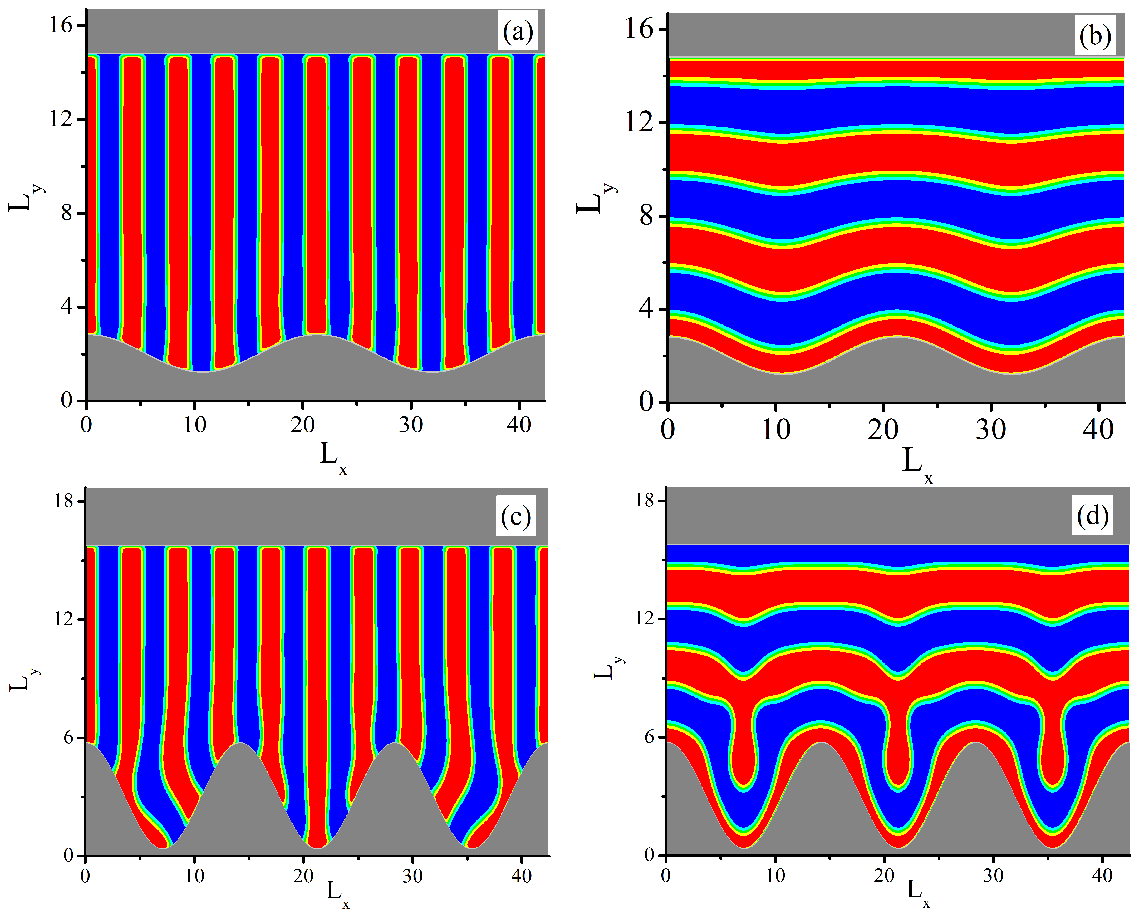}}
\caption{
\textsf{SCFT calculation of BCP lamellae in contact with a moderate corrugated substrate in (a) and (b), $q_sR\simeq 0.25$,  and with a pronounced corrugated substrate  in (c) and (d), $q_sR\simeq 1.23$. For (a) and (b), the corrugation wavenumber is $q_s=2(2\pi/L_x)$ and its amplitude $R=0.85$.
The perpendicular lamellar phase (L$_\perp$) in (a) is almost unperturbed compared to its bulk phase, while the parallel one (L$_\parallel$) in (b) closely follows the surface contour. For (c) and (d), $q_s=3(2\pi/L_x)$ and $R= 2.76$. Both L$_\perp$ in (c) and L$_\parallel$ in (d) are strongly deformed.
In all figure parts, calculations are done in the strong segregation limit, $N\chi_{AB}=25$, resulting in a natural lamellar periodicity, $L_0=4.25$. The surface preference is $u=3.05$ for the bottom substrate and $u_{\rm top}=0$ for the top surface. The color code are the same as those in Fig.~2. The lateral box size is $L_x=42.5$, the average film thickness $L=12.75$. All lengths are rescaled by $R_g$ in all figures.
}}
\end{center}
\end{figure*}

When the BCP film thickness, $L$, differs from an integer multiple of $L_0$, the BCP chains have to stretched
or compressed, as the total film volume is incompressible and space-filling (in our formalism, $\zeta=1000$ is large enough and model an incompressible system). In order to minimize such a
confinement effect and focus mainly on how surface roughness affects the lamellar orientation, we adjust the
box size such that the film thickness
$L$ corresponds to a local minimum of the L$_{\parallel}$ free energy. This
is done by investigating the free energy difference between parallel and perpendicular lamellae, $\Delta
F=F_{\parallel}-F_{\perp}$, where $F_{\parallel}$ and $F_{\perp}$ are, respectively, the lamellar
free-energies of the two orientations.

\begin{figure*}[h!t]
\begin{center}
{\includegraphics[bb=0 0 337 251, scale=0.7,draft=false]{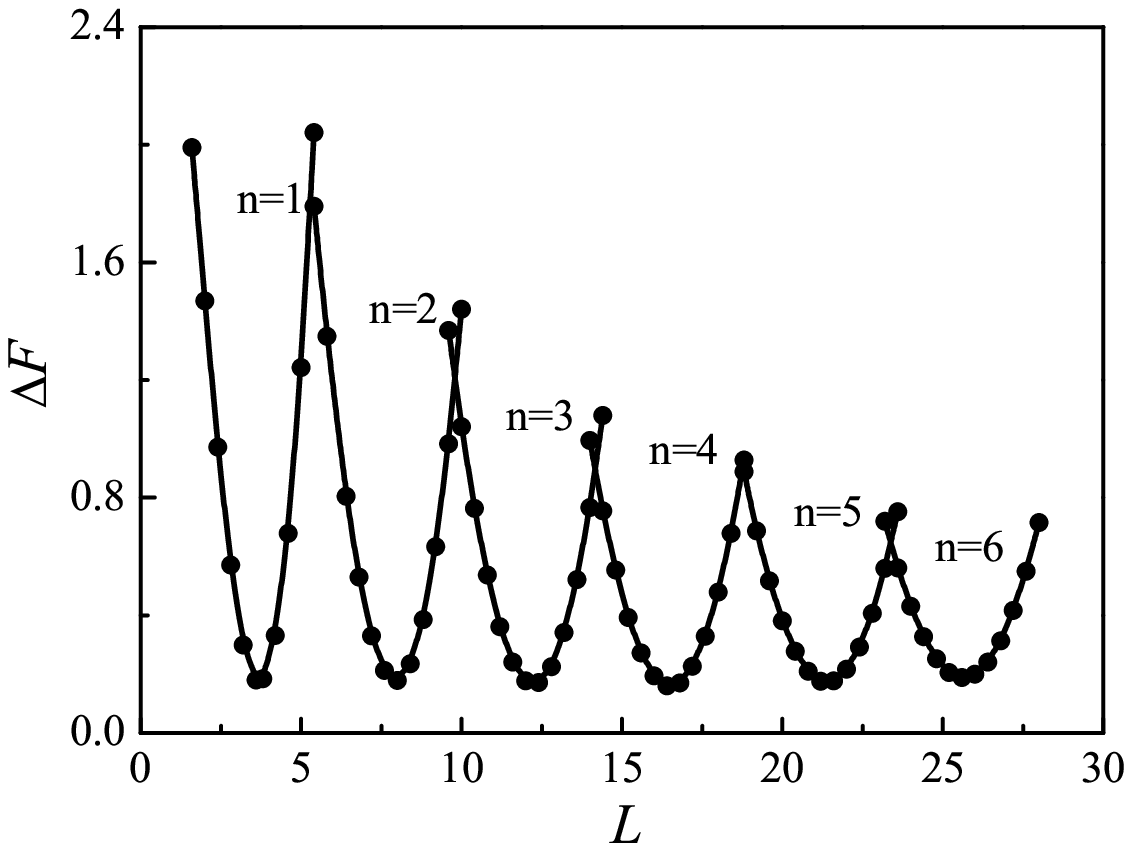}}
\caption{
\textsf{The free-energy difference between parallel and perpendicular BCP lamellar orientations, $\Delta
F=F_{\parallel}-F_{\perp}$, in units of $n_ck_BT/L$, as function of the film thickness, $L$, where $n_c$ is the
number of chains and $k_BT$ is the thermal energy. The parallel free-energy, $F_{\parallel}$, is calculated separately for
$n=1,2,\dots, 6$ parallel lamellae confined between two flat neutral surfaces. The perpendicular free energy,
$F_{\perp}$, corresponds to three periods of perfect perpendicular lamellae. The other parameters are
$L_x=12.75$ and $L_0=4.25$ (or equivalently, $N\chi_{\rm{AB}}=25$).}
}
\end{center}
\end{figure*}\label{fig3}

Figure~4 shows the dependence of $\Delta F$ on $L$ with $N\chi_{\rm{AB}}=25$. We repeat the calculation of $\Delta F$ for various lamellar layers, $n=1,2,\dots,6$, between two flat and neutral surfaces. Clearly, the rescaled $\Delta F$ has always a local minimum when $L$ equals to an integer number times the natural periodicity, $L=nL_0$, in accord with previous SCFT calculations of Takahashi {\it et al}~\cite{Takahashi12}. It is known that $L_0$ depends on the value of $N\chi_{\rm{AB}}$. Therefore, for different values of $N\chi_{\rm{AB}}$ we have to repeat this calculation and obtain the free energy in order to find the appropriate local minima, as in Fig.~4.

It is important to note that when a thin BCP film is confined between two surfaces, $L_0$ is also a function of the film thickness, $L$, and differs from its bulk value \cite{Takahashi12}. Moreover, the free energy of confined BCP films also depends on the surface preference field, $u$ \cite{Matsen97,Geisinger99, Man10,Tsori01}. Therefore, the parameters $N\chi_{\rm{AB}}$, $L$ and $u$ play an important role in determining whether the equilibrium orientation will be parallel (L$_{\parallel}$) or perpendicular (L$_{\perp}$), as will be further presented below.

\begin{figure*}[h!t]
\begin{center}
{\includegraphics[bb=0 0 332 257, scale=0.7,draft=false]{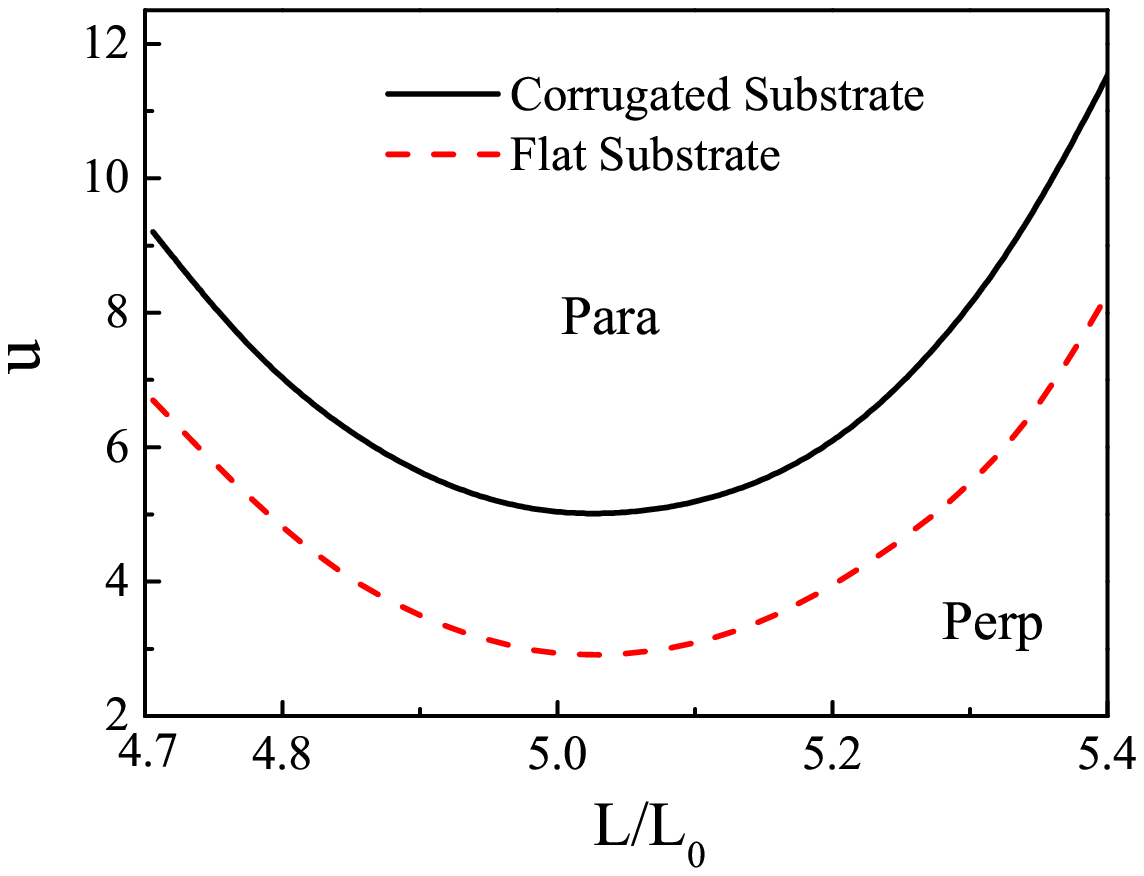}}
\caption{
\textsf{The L$_{\perp}$ -- L$_{\parallel}$ phase diagram in terms of the rescaled film thickness $L/L_0$ and
substrate preference $u$, for a corrugated substrate (solid  line), and for a flat substrate (dashed line).
The top surface is flat and neutral ($u_{\rm{top}}=0$). The lines separate between a stable L$_{\perp}$
(perp) phase below and L$_{\parallel}$ (para) phase above. The corrugated
substrate is characterized by $q_s=9(2\pi/L_x)\simeq 1.33$ and $R=0.45$,  yielding a roughness parameter, $q_sR\simeq 0.6$. Other parameters are $L_x=42.5$ and $L_0=4.25$ (or equivalently, $N\chi_{\rm{AB}}=25$).}
}
\end{center}
\end{figure*}

We first  compute the perp-to-para (L$_\perp$--L$_\parallel$) phase diagram (shown in Fig.~5) in terms of the film
rescaled thickness, $L/L_0$, and bottom surface preference, $u$, for a corrugated substrate. For comparison,
the calculation is then repeated for another system having a flat bottom surface, with the same parameters: $L_x=42.5$, $u_{\rm{top}}=0$ and $N\chi_{\rm{AB}}=25$. The film thickness varies around $5L_0$: $4.7 \leq L/L_0\leq 5.4$. The wavenumber of the corrugated substrate is $q_s=9(2\pi/L_x)$, while the corrugation magnitude is fixed,
$R=0.45$, resulting a substrate roughness $q_sR\simeq 0.60$. The phase diagram is obtained by starting with an initial condition of either an L$_{\perp}$ phase of ten periods or an L$_{\parallel}$ of five periods. After numerical convergence, the corresponding free energies are compared. The results show that the rough substrate greatly affects the phase diagram as compared with flat substrate. The L$_{\perp}$ -- L$_{\parallel}$ phase-transition line for the corrugated case is shifted upward, which means that the L$_{\perp}$ phase has a larger stability range for rough substrates than for flat ones.
This conclusion qualitatively agrees with previous analytical and experimental studies~\cite{Yoav05,Sivaniah05,Char13}. It shows that rough substrates, just like chemical-patterned substrates and nano-Imprint surfaces~\cite{Man10}, can enhance the stability of the L$_{\perp}$ phase as compared with a flat
substrate with the same surface preference field, $u$.

\subsection{The Substrate Effect on Para-to-Perp Transition} 

\begin{figure*}[h!t]
\begin{center}
{\includegraphics[bb=0 0 350 253, scale=0.7,draft=false]{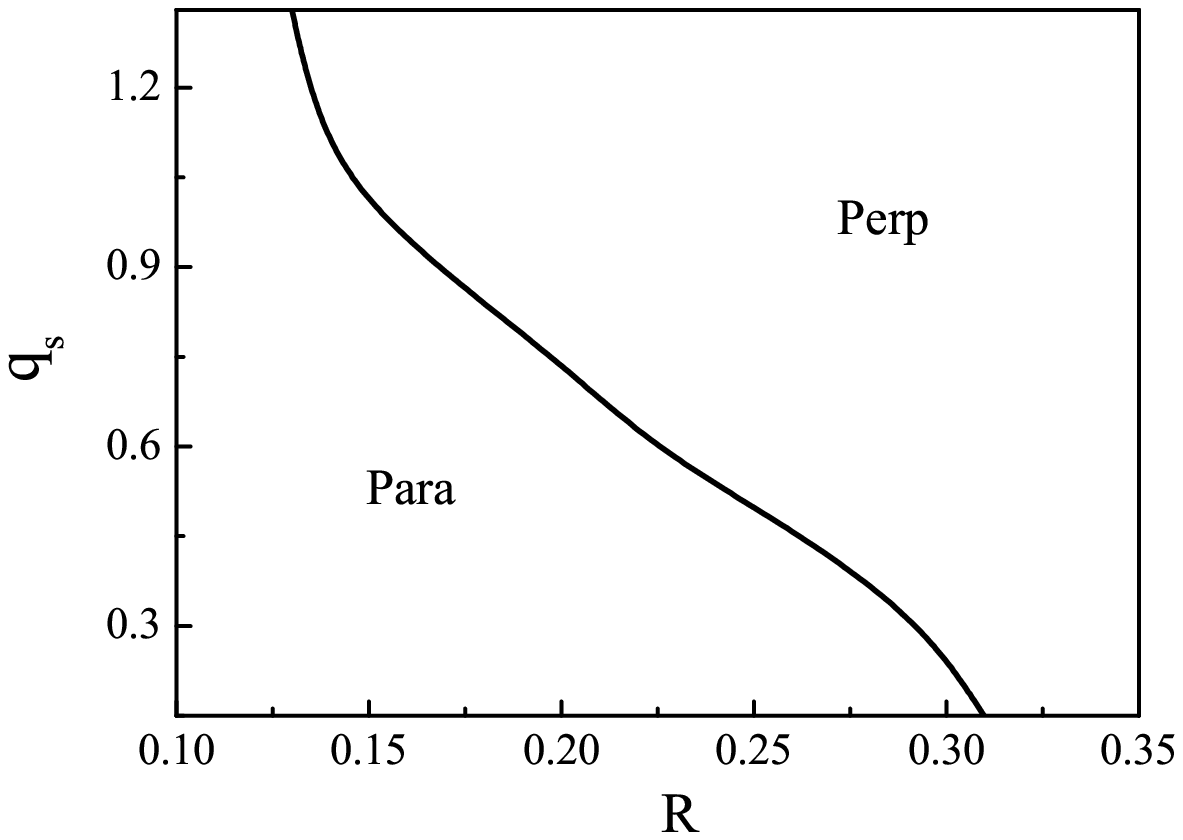}}
\caption{\textsf{The L$_{\perp}$--L$_{\parallel}$ phase diagram in terms of $R$ and $q_s$.
The parameters used are: $L=21.25$, $L_x=2L=42.5$, $L_0=4.25$ (or equivalently, $N\chi_{\rm{AB}}=25$), $u=3.05$, and $u_{\rm{top}}=0$.}}
\end{center}
\end{figure*}

We proceed by studying quantitatively the effect of substrate roughness on the relative stability of the
L$_{\perp}$ and L$_{\parallel}$ phases. We focus on the role played by i)~the substrate geometry, including
lateral wavenumber, $q_s=2\pi/L_s$, and roughness amplitude, $R$, ii)~the relative surface preference towards
the two BCP components, $u$, and iii)~BCP film properties, including film thickness, $L$, the Flory-Huggins
parameter, $N\chi_{AB}$, as well as the BCP natural periodicity, on the L$_\parallel$--L$_\perp$ phase-transition.

Figure~6 shows the L$_\parallel$--L$_\perp$ phase diagram in terms of the corrugation parameters, $R$ and
$q_s$, for a fixed bottom substrate preference ($u=3.05$) and a flat neutral top surface ($u_{\rm top}=0$).
For fixed $q_s$, while increasing $R$, an L$_{\parallel}$-to-L$_{\perp}$ phase-transition is reached because the elastic deformation of the L$_{\parallel}$ lamellae along the corrugated surface becomes bigger, favoring the L$_{\perp}$ orientation. For fixed value of $R$, an increase of $q_s$ also induces an L$_{\parallel}$-to-L$_{\perp}$ phase-transition. This can be understood as smaller $q_s$ (while keeping $R$ constant) means that the substrate effectively is flatter, and the L$_{\parallel}$ lamellae are losing less elastic deformation energy. In the limit of $q_s\rightarrow0$ (corresponding to an unrealizable large simulation box),
the substrate approaches to a flat substrate. Then, the L$_{\parallel}$ phase will be more stable than
L$_{\perp}$ because of the substrate preference towards one of the two BCP components.

\begin{figure*}[h!t]
\begin{center}
{\includegraphics[bb=0 0 339 251, scale=0.7,draft=false]{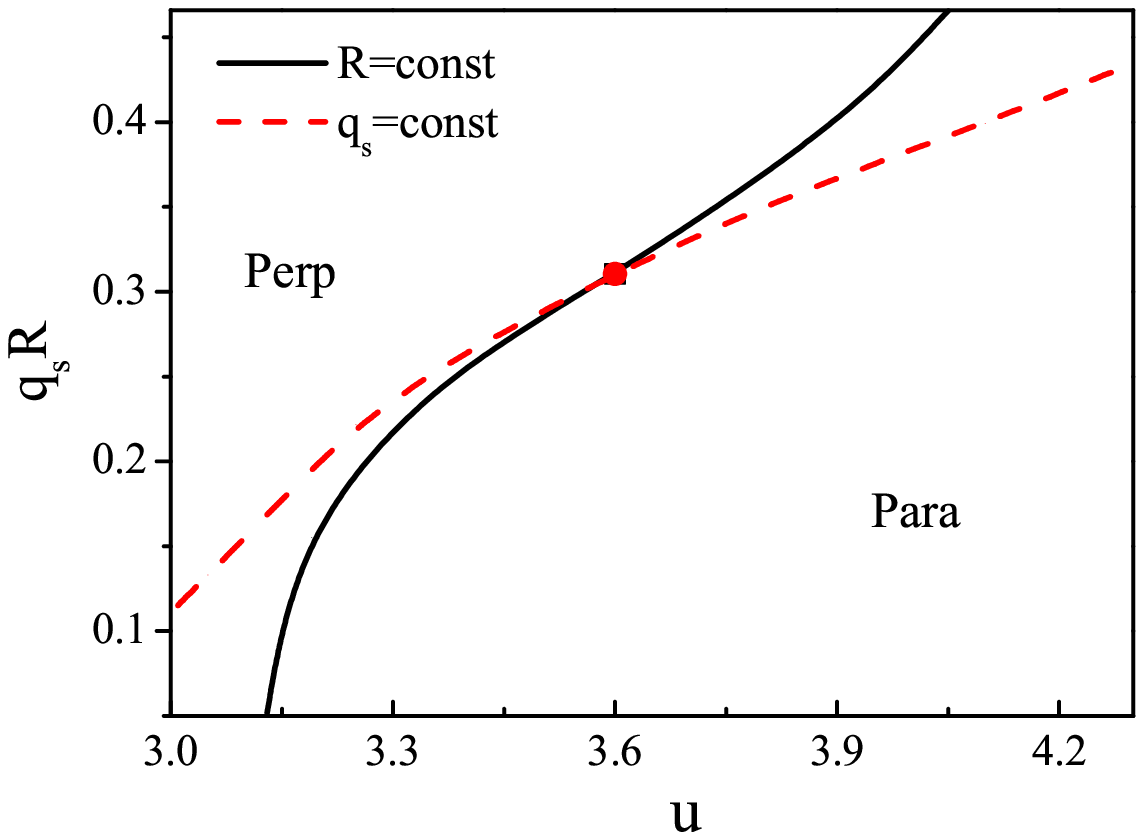}}
\caption{
\textsf{Comparison between two ways (solid and dashed lines) to obtain the L$_\parallel$-to-L$_\perp$
phase-transition plotted in the ($q_sR$, $u$) plane.
The solid line is calculated for constant $R=0.35$. The roughness wavenumber $q_s=n_s(2\pi/L_x)$ is varied discretely,
where $n_s=1,2,3,\dots,9$ is
the number of periods within the lateral box width. The dashed line corresponds to constant
$q_s=6(2\pi/L_x)$, while $R$ varies from $0.1$ to $0.5$.  The
results show that $(q_sR)^*$ at the phase transition increases faster by varying $q_s$ than by varying $R$. All other parameters are the same as in Fig.~6.}}
\end{center}
\end{figure*}

As discussed in the section III.A above, the surface preference, $u$, is an important factor in determining the final lamellar orientation. Moreover, $q_sR$ is usually used to parameterize the substrate roughness in experiments. We investigate the effects of the substrate $u$ on the critical value, $(q_sR)^*$, corresponding to the L$_{\parallel}$--L$_{\perp}$ phase-transition, while keeping the top surface neutral ($u_{\rm{top}}=0$) and flat (see Fig.~7). Our results show that $(q_sR)^*$ increases as function of $u$. This means that for larger $u$, larger substrate roughness is needed to induce an L$_{\parallel}$-to-L$_{\perp}$ phase-transition. This is due to the competition between the energy cost of elastic deformation and gain in surface energy. Increasing $q_sR$ leads to more elastic deformation in the L$_{\parallel}$ phase as compared with L$_{\perp}$, resulting in a phase transition from an L$_{\parallel}$ to an L$_{\perp}$ phase. Oppositely, increasing $u$ makes the L$_{\parallel}$ phase more stable because of the gain in surface energy, leading to a phase transition from L$_{\perp}$ to L$_{\parallel}$.

\begin{table*}[h!t]
\caption\textsf{The critical substrate roughness for various rescaled BCP film thicknesses}
\begin{center}
\begin{tabular*}{0.75\textwidth}{@{\extracolsep{\fill}}cccccc}
\hline
\hline
$L/L_0$ & 2 & 3 & 4 & 5 & 6\\
\hline
$(q_{s}R)^{*}$ &0.15&0.17&0.18&0.21&0.26\\
\hline
\hline
\end{tabular*}
\end{center}
\end{table*}

We present in Fig.~7 two ways to change the values of $q_sR$. First, we fix the corrugation
amplitude, $R=0.35$, and change the wavenumber discretely by taking these values as $q_s=n_s(2\pi/L_x)$, where $n_s=1,2,\dots,9$ is the number of lateral periods of the substrate. Next, we fix $q_s=6(2\pi/L_x)$ and change $R$ from $0.1$ to $0.5$. The simulation box size is set to $L_x=42.5$, $L=21.25$, and $N\chi_{AB}=25$. For the calculations, L$_{\perp}$ has ten BCP lamellar periods as its initial condition while L$_{\parallel}$ has five. After convergence, the L$_{\perp}$ and L$_{\parallel}$ free energies are compared, and clearly, for both phases, $(q_sR)^*$ is an increasing function of $u$. However, for the same value of $u$, the value of $(q_sR)^*$ is different for constant $R$ and constant $q_s$, with a $(q_sR)^*$ difference of the order of $10^{-2}$. This indicates that $q_sR$ is not exactly a scaling field for the L$_{\parallel}$--L$_{\perp}$ phase-transition. The reason is that the contour length of the substrate has a small difference (also on the order of $10^{-2}$) for different combinations of $q_s$ and $R$, while keeping their product $q_sR$ the same. Such differences will further result in a different elastic deformation and surface energies of the film.

\subsection{The Film Effect on the Para-to-Perp Phase Transition} 

Another important parameter that affects $(q_sR)^*$ is the BCP film thickness, $L$. Table~I presents the
dependence of $(q_sR)^*$ on $L$, obtained with fixed preference of the bottom and top surfaces, $u=3.05$ and
$u_{\rm{top}}=0.15$, respectively. We add a small top surface preference, $u_{\rm top}>0$, to mimic the experiments, where BCPs usually have non-zero preference in their surface interaction with the air. More discussion about the top surface preference effects on the L$_{\parallel}$--L$_{\perp}$ phase-transition will be addressed in the section IV. As mentioned in Fig.~4, we choose the film thickness $L$ to be an integer multiple of the L$_{\parallel}$ periodicity, i.e. $L=nL_0$, where $n=2,3,\dots, 6$, in order to minimize the film confinement effects. The L$_{\parallel}$ free energy is compared with the L$_{\perp}$ one with $L_x=2L_0$, while all other parameters are kept the same. The numerical results indicate that $(q_sR)^{*}$ increases for thicker BCP film. This is understandable because the elastic deformation induced by the corrugated substrate is a surface effect with limited penetration into the BCP film. Therefore, larger $q_sR$ values are needed to induce the L$_{\parallel}$-to-L$_{\perp}$ phase-transition for thicker films.

Finally, we examine the effect of $N\chi_{\rm{AB}}$ (or $L_0$) on the L$_{\parallel}$--L$_{\perp}$ phase-transition in Fig.~8. As mentioned above, varying $N\chi_{\rm{AB}}$ will also change the BCP lamellar periodicity, $L_0$. In order to stay in the minimal confinement free-energy, we set $L=nL_0$, where $L_0$ varies for different $N\chi_{\rm{AB}}$. Because the minimum in $F_\perp$ corresponds to $L_x=nL_0$, we obtain the value of $L_0$ for different $N\chi_{\rm{AB}}$ by examining the dependence of $F_{\perp}$ on $L_x$, while keeping $L$ fixed. After
determining the values of $L_0$, we can adjust our simulation box accordingly. The $L_0$ values for different $N\chi_{\rm{AB}}$ are shown in Table~II, where it can be seen that $L_0$ is an increasing function of $N\chi_{\rm{AB}}$~\cite{Sivaniah08}. For all calculations in Fig.~8, the parameters are: $L_x=2L_0$, $u=3.05$ and $u_{\rm{top}}=0.15$. The $L$ values for different $N\chi_{\rm{AB}}$ are set to fulfill $L=3L_0$ (see Table~II).

From Fig.~8(a), one sees that $(q_sR)^*$ decreases as $N\chi_{\rm{AB}}$ increases. This can be understood because
when $N\chi_{\rm{AB}}$ increases, the lamellar periodicity, $L_0$, also increases, and the local deformation
of BCP lamellae becomes larger. For example, in the limit of $q_s/q_0\rightarrow0$, the corrugated substrate is equivalent to a flat one, and the BCP lamellae will not be deformed anymore, resulting in no elastic deformation. In other words, the energy cost of the elastic deformation increases as $q_s/q_0$ increases. Then, smaller values of $(q_sR)^*$ are needed to induce the L$_{\parallel}$-to-L$_{\perp}$ phase-transition. The tendency of $(q_sR)^*$ to decrease when $L_0$ increases is shown in Fig.~8(b) and is consistent with the results in Table~I. When $L_0$ increases, it means that the value of $L/L_0$ in Table~I decreases, and the same tendency of $(q_sR)^*$ on $L_0$ is obtained.

\begin{table*}[h!t]
\caption\textsf{Lamellar periodicity, $L_0$, and average film thickness, $L=3L_0$, for various $N\chi_{\rm{AB}}$
values.}
\begin{center}
\begin{tabular*}{0.75\textwidth}{@{\extracolsep{\fill}}cccccc}
\hline
\hline
$N\chi_{\rm{AB}}$ & 18 & 20 & 22 & 25 & 28\\
\hline
$L_0$ & 3.90 & 4.00 & 4.10 & 4.25 & 4.35\\
$L$ & 11.7 & 12.00 & 12.30 & 12.75 & 13.05\\
\hline
\hline
\end{tabular*}
\end{center}
\end{table*}

\begin{figure*}[h!t]
\begin{center}
{\includegraphics[bb=0 0 341 249, scale=0.65,draft=false]{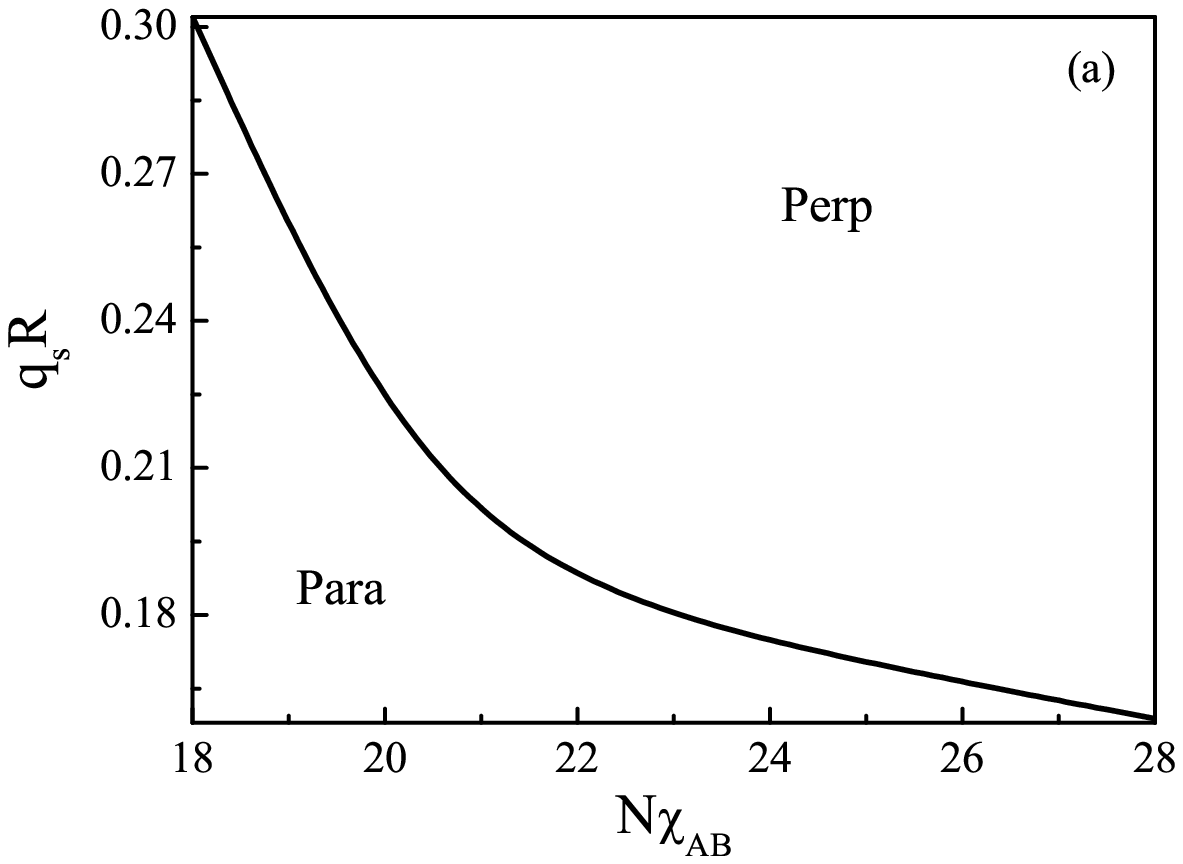}}
{\includegraphics[bb=0 0 341 253, scale=0.65,draft=false]{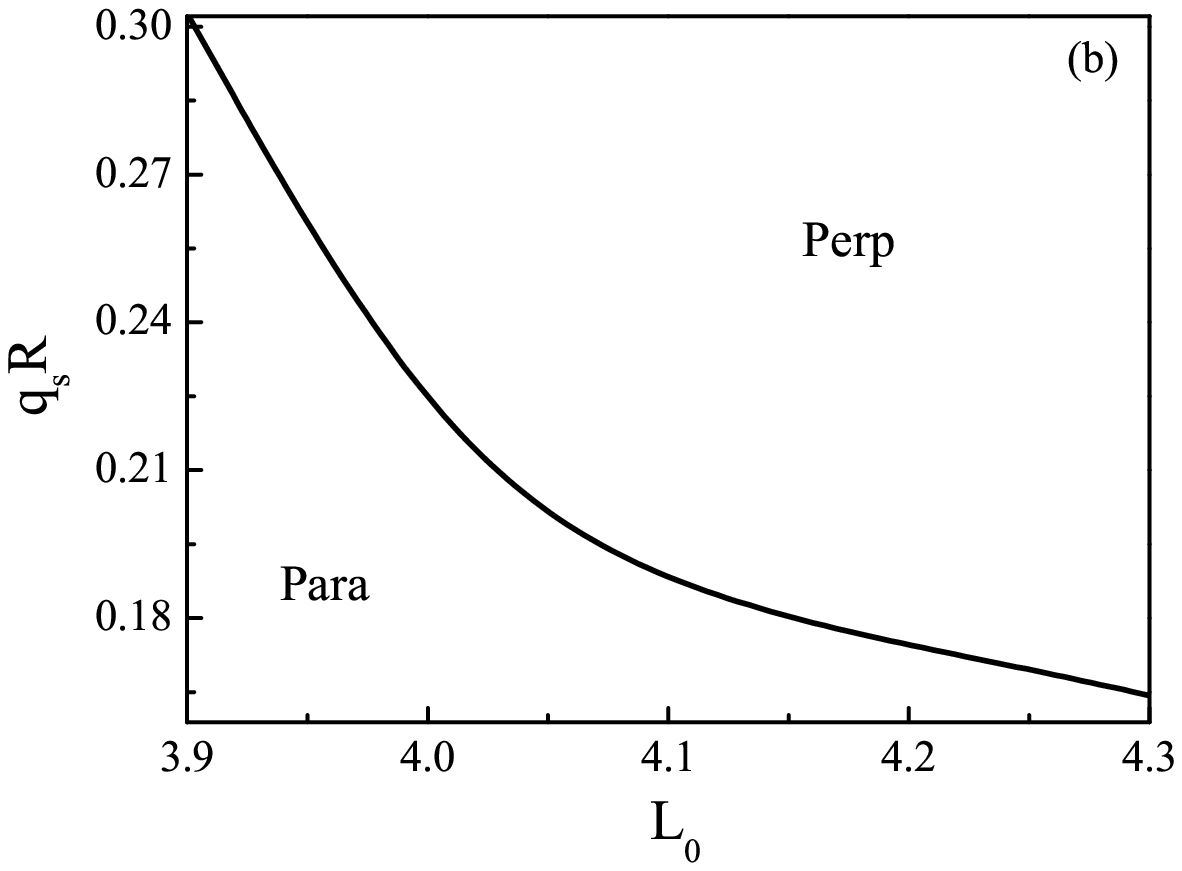}}
\caption{
\textsf{(a) The L$_{\perp}$ -- L$_{\parallel}$ phase diagram in terms of $q_sR$ and the interaction between A
and B blocks, $N\chi_{\rm{AB}}$. The box size is adjusted to have two periods of L$_{\perp}$ and three
periods of L$_{\parallel}$ for various values of $N\chi_{\rm{AB}}$. (b) The equivalent L$_{\perp}$ -- L$_{\parallel}$ phase diagram expressed in terms of $L_0$ and
$q_sR$. Other parameters are: $u=3.05$,
$u_{\rm{top}}=0.15$ and $q_s=2\pi/L_x$.}}
\end{center}
\end{figure*}

\section{Discussion} 

\subsection{Comparison with Experiments} 

Our study has been motivated by the experimental works of Sivaniah \emph{et al} \cite{Sivaniah05} and Char's
group \cite{Char13}. In Ref. 16, PS-b-PMMA films are casted onto an array of polyimide (PIM) substrates having
different roughness. The study found the critical roughness of the PIM substrate, $(q_sR)^*$, separating
the stable L$_\perp$ lamellae for $q_sR>(q_sR)^*$ from the region where the L$_\parallel$ lamellae are more
stable, $q_sR<(q_sR)^*$. In a more recent work by Char's group \cite{Char13}, both lamellar and cylindrical phases of PS-b-PMMA were spin-coated onto two different substrates covered with an ordered nanoparticle (NP) monolayer. In the first set-up, the NP size is $R=6\,\rm{nm}$ and its repeat period is $q_s\simeq 0.75\,\rm{nm}^{-1}$, resulting in $q_sR\simeq 4.5$. For the second set-up, the NP size is $R=22\,\rm{nm}$, the repeat period is $q_s\simeq 0.26\,\rm{nm}^{-1}$, which leads to $q_sR\simeq 5.72$. For both lamellae and cylinders, $L_\parallel$ orientation was obtained in the first case $(q_sR\simeq 4.5)$, and $L_\perp$ orientation for the second one $(q_sR\simeq 5.72)$. The conclusion from these experiments is that increasing the substrate roughness can induce an L$_{\parallel}$-to-L$_{\perp}$ phase-transition. All of our results nicely support these findings, although we can justify less our SCFT results in a quantitative manner for $q_s R>1$.

Sivaniah \emph{et al} \cite{Sivaniah05} have also shown that the value of $(q_sR)^*$ varies with the BCP molecular
weight. Keeping all experimental conditions the same, the value of $(q_sR)^*$ for 38K-38K
PS-b-PMMA was found to be $0.37\pm0.02$ and for 18K-18K is it $0.41\pm0.02$. Furthermore,
the natural periodicity for the 38K-38K system is $L_0=36.7\,\rm{nm}$, and for 18K-18K it is
$L_0=28.6\,\rm{nm}$. Because the experiments were conducted for only two BCP chain lengths, it was not possible to infer any trend of lower $(q_sR)^*$ values for higher molecular weights. However, our results indicate such a trend, where the values of $L_0$
increase from $3.90$ to $4.35$, and corresponds to a decrease of $(q_sR)^*$.

\begin{figure*}[h!t]
\begin{center}
{\includegraphics[bb=0 5 347 280, scale=0.7,draft=false]{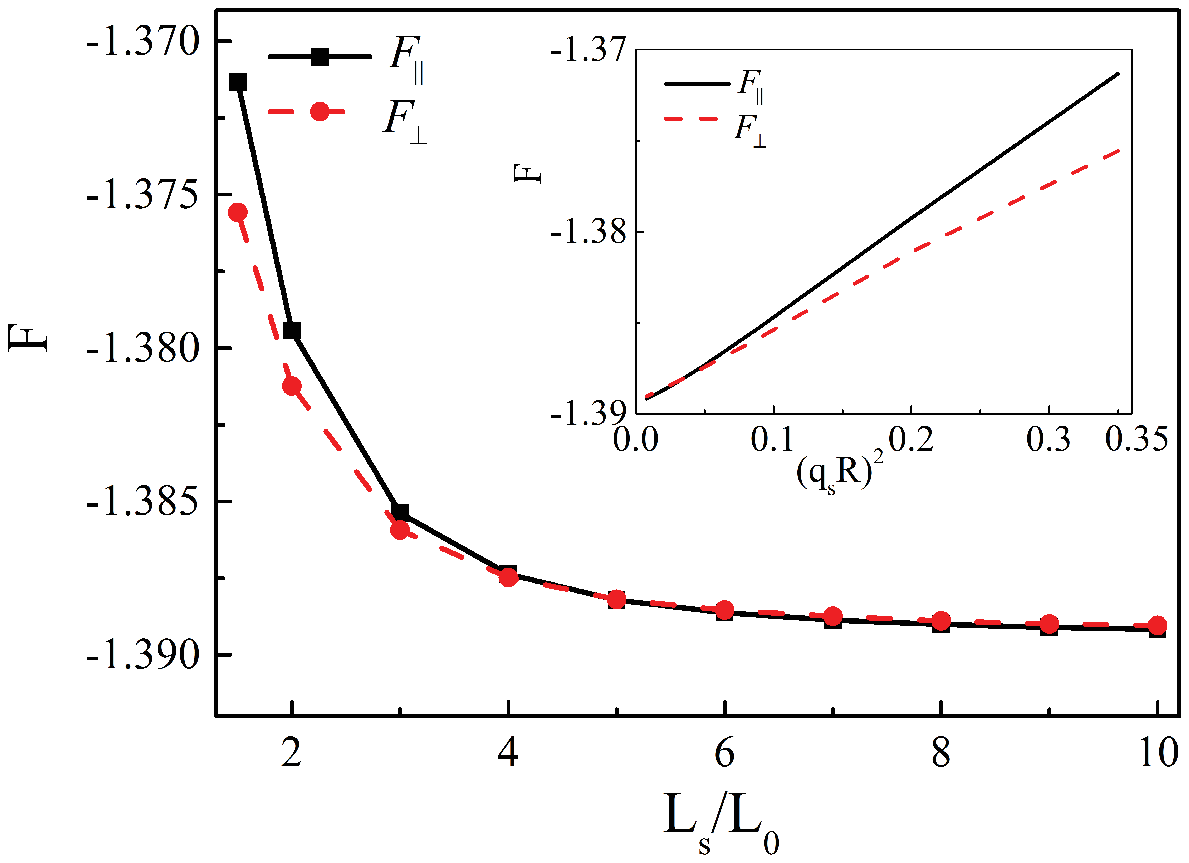}}
\caption{
\textsf{The dependence of $F_{\parallel} $ and $F_{\perp}$ on the lateral wavelength, $L_s=2\pi/q_s$.
The value of
$L_s$ varies from $1.5L_0$ to $10L_0$ with $L_0=4.25$.
In the inset, the two free energies are shown to scale, within a good approximation, with $(q_sR)^2$.
Other parameters are: $L_x=42.5$, $L=21.25$, $R=0.6$, $u=3.3$,
$u_{\rm{top}}=0$, and $N\chi_{\rm{AB}}=25$.}}
\end{center}
\end{figure*}

\subsection{Comparison with Previous Models} 

\begin{figure*}[h!t]
\begin{center}
{\includegraphics[bb=0 0 350 300, scale=0.7,draft=false]{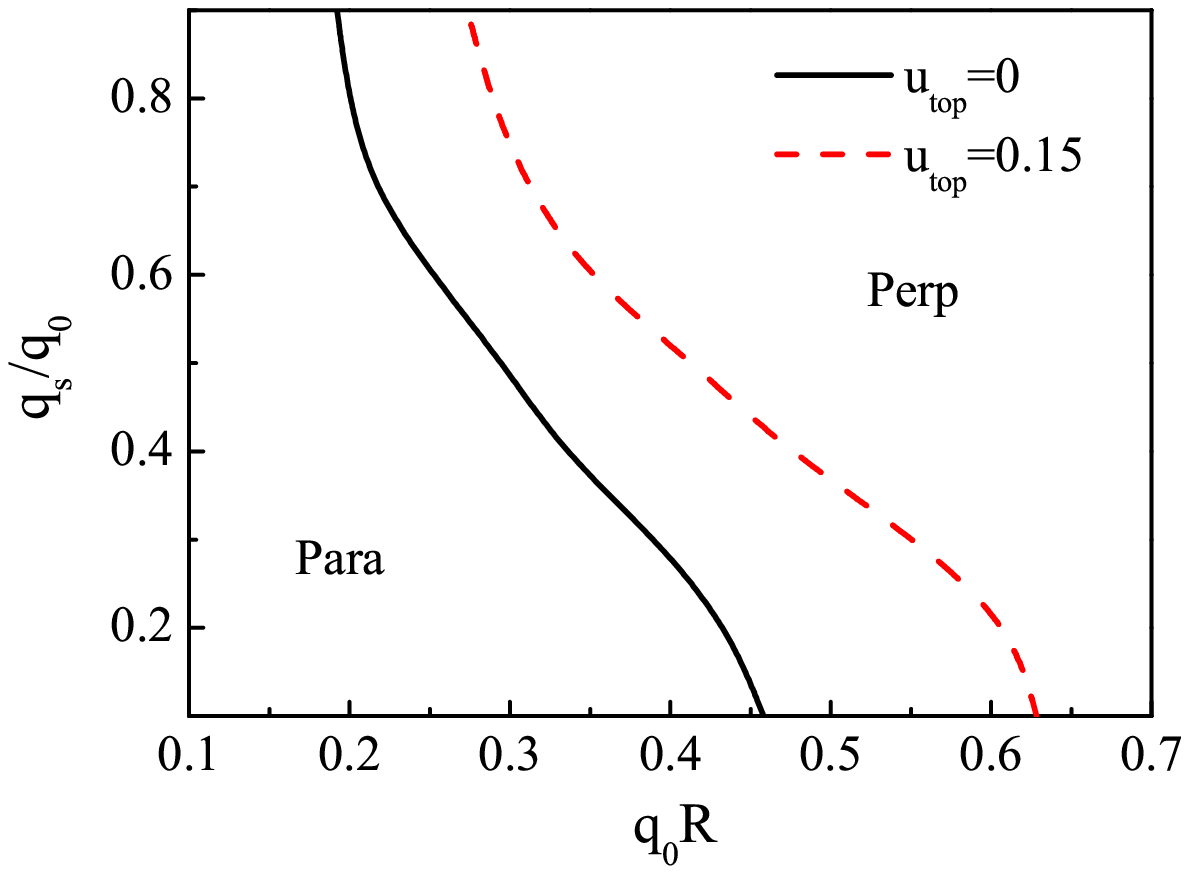}}
\caption{
\textsf{Comparison of the L$_\perp$ -- L$_\parallel$ phase diagram in the ($q_0R$, $q_s/q_0$) plane, between a neutral top surface $u_{\rm top}=0$ (solid black line) and a non-neutral one, $u_{\rm top}=0.15$ (dashed red line). All other parameter values are
as in Fig.~6.}
}
\end{center}
\end{figure*}

In a closely-related analytical study, Tsori \emph{et al} \cite{Yoav03,Yoav05} used the analogy between smectic
liquid crystals and lamellar BCP, and compared the phenomenological free-energies of L$_{\parallel}$ and
L$_{\perp}$ phases on corrugated substrate, and study their relative stability. Their results can be summarized
by the following three findings. While keeping all other parameters fixed, these authors found that {\it i)}~increasing $R$ induces an L$_{\parallel}$-to-L$_{\perp}$ phase-transition; {\it ii)}~increasing $q_s$ results in an L$_{\perp}$-to-L$_{\parallel}$ phase-transition; and, {\it iii)}~increasing $q_0$ leads to an L$_{\parallel}$-to-L$_{\perp}$ phase-transition. Our findings agree with the first finding of Tsori \emph{et al},
as well as with the experimental findings discussed in the section IV.A above. Increasing the value of $R$ (while all other parameters are fixed) means larger substrate roughness, and therefore, it induces an L$_{\parallel}$ -- L$_{\perp}$ phase-transition. However, the second and third findings of Tsori \emph{et al} are contrary to experimental trends~\cite{Sivaniah05}, as well as to our SCFT calculations.

One of the results of Tsori \emph{et al} \cite{Yoav03,Yoav05} is that $F_{\parallel}$ is an increasing function of the lateral wavelength, $F_{\parallel}\sim (R/q_s)^{2}\sim q_s^{-2}$, while $F_{\perp}$ scales as $(q_sR)^2$. However, our SCFT results show an opposite trend for $F_{\parallel}\sim (q_sR)^2\sim q_s^2$, while the same scaling for $F_{\perp}\sim (q_sR)^2$. This behavior of the two free energies is shown in Fig.~9, and in the inset the scaling with $(q_sR)^2$ can be clearly seen. We note that in an even earlier work by Turner and Joanny~\cite{Turner1992}, the free energy of both $F_{\parallel}$ and $F_{\perp}$ stack on a corrugated substrate were found to scale as $(q_sR)^2$ (while all other parameters are fixed).
We conjecture that the discrepancy between our SCFT and the previous analytical results~\cite{Turner1992,Yoav03,Yoav05} is due mainly to two reasons. One, is the fact that our stack has a finite width and the top surface is affecting the penetration length, and hence the delicate balance between the two orientations. The second reason is related to local deformations of BCP chains close to the corrugated substrate. These deformations are not well described within models that draw on the analogy with a continuum theory of smectic liquid crystals, and which assumes small and gradual deformations.

\subsection{Non-neutral Air/Polymer Interface} 

It is known that when the top surface (polymer film/air) has a preference toward one of the two BCP blocks,
it can induce an L$_{\parallel}$ orientation inside the film. In many applications (e.g., in nanolithography),
it is highly desired to circumvent the $L_{\parallel}$ phase by using the so-called `top coats'
\cite{Bates12,Yoshida13,Son14}. Moreover, Khanna \emph{et al} \cite{Khanna06} in experiments and Matsen
\cite{Matsen10} in theory have shown that the BCP architecture can also affect the polymer/air surface tension,
and facilitates the formation of L$_{\perp}$ lamellae.

Figure~10 shows a comparison of the L$_{\parallel}$--L$_{\perp}$ phase diagram in terms of $R$ and $q_s$
between a system with a top surface field, $u_{\rm{top}}=0.15$, and a second system with a neutral top surface,
$u_{\rm{top}}=0$ (same as in Fig.~5). From Fig.~10 it can be seen to what degree
the non-zero top surface affects the
relatively stability of the L$_{\perp}$ and L$_{\parallel}$ phases.
When a small surface preference is added to the top surface,
the L$_{\parallel}$--L$_{\perp}$ phase-transition line shifts to the right (the red dashed line); namely,
the transition value, $(q_sR)^*$ increases.
Our study is instrumental as it shows a possible solution utilizing a rough substrate to overcome the parallel
orientation induced by the commonly-found air/film preference, $u_{\rm{top}}\neq 0$. Therefore, combining the
`top coats', varying BCP architecture, as well as employing rough substrates may offer an effective way to obtain
perpendicular orientation of BCP nano-structures, even in cases where BCPs have significant different surface
tension between the air and the two blocks of the BCP.

\section{Conclusions} 

In this article, we address the influence of a non-flat substrate on the relative stability between
the two orientations, L$_{\parallel}$ and L$_{\perp}$,  of lamellar BCP phases. The thin lamellar film is confined between a top flat surface and bottom corrugated substrate of a shape, $R\cos(q_sx)$, with a single $q$-mode of lateral undulations. The competition between the energy cost of elastic deformation and gain in surface energy of BCP lamellae results in an L$_{\parallel}$--L$_{\perp}$ phase-transition.

We comprehensively and systematically studied the combined effect of the rough substrate with lateral wavenumber, $q_s$, and magnitude $R$, as well as the interface energy between the BCP and the surface, film thickness and the Flory parameter on the L$_{\parallel}$-to-L$_{\perp}$ phase-transition. Our results show that  increasing the substrate roughness, $q_sR$, induces an L$_{\parallel}$-to-L$_{\perp}$ phase-transition. Moreover, the critical value of the substrate roughness, $(q_sR)^*$, corresponding to the L$_{\parallel}$--L$_{\perp}$ phase-transition, increases as the surface preference towards one of the two blocks, $u$, increases, or as the film thickness becomes thicker.
On the other hand, it decreases when the Flory parameter, $N\chi_{\rm{AB}}$, or the natural periodicity, $L_0$, increases.

We focused in this study on a few key factors that enhance or induce the phase-transition from L$_\parallel$ into the L$_{\perp}$ phase, as is desired in applications. As detailed in the section~IV, our predictions are consistent with several experimental findings. Furthermore, as our study is systematic, its predictions can be further tested experimentally by using di-BCP with different chain lengths and periodicities, and by changing in a tunable fashion the corrugated substrate, the top surface preference as well as the film thickness, in order to determine the optimal condition for the enhance stability of the L$_\perp$ phase.

In addition to the effect of substrate roughness on the parallel-to-perpendicular orientation transition,
corrugated substrates have been shown to improve in-plane ordering of thin BCP films~\cite{Park09b,Park09a,Vega13}. We hope that in the future more detailed 3d calculations will shed more light on various possibilities that non-flat substrates may improve the in-plane ordering and defect annihilation of BCP lamellar phases.

\bigskip
{\bf Acknowledgement.}~~
This work was supported in part by grants 21404003, 21434001 and 21374011 of the National Natural Science Foundation of China (NSFC), the 973 program of the Ministry of Science and Technology (MOST) 2011CB808502, and the Fundamental Research Funds for the Central Universities. We thank Henri Orland and  Yoav Tsori for useful
discussions and comments. D.A. acknowledges support from the Israel Science Foundation (ISF) under Grant No.\ 438/12 and the United States--Israel Binational Science Foundation (BSF) under Grant No.\ 2012/060.

\newpage

%

\end{document}